\documentstyle [aps,prl,epsf] {revtex}

\begin{document}

\renewcommand{\[}{\begin{equation}}
\renewcommand{\]}{\end{equation}}

\title{Critical Current in the High-$T_c$ Glass model}
\author{Ingo Morgenstern, Werner Fettes, and Thomas Husslein}
\date{\today}

\maketitle

\begin{abstract}
\begin{quote}
\parbox{16cm}
{\small 
The high-$T_c$ glass model can be combined with the repulsive tt'--Hubbard model 
as microscopic description of the striped domains found in the high-$T_c$ 
materials. In this picture the finite Hubbard clusters are the origin of 
the d-wave pairing.
In this paper we show, that the glass model can also explain the  
critical currents usually observed in the high-$T_c$ materials. 
We use two different approaches to
calculate the critical current densities of the high-$T_c$ glass
model. Both lead to a strongly anisotropic critical current. Finally we give an 
explanation, why we expect nonetheless a nearly perfect isotropic critical current 
in the high-$T_c$ superconductors.
}
\end{quote}
\end{abstract}

\bigskip

{{\bf Keywords:} High-$T_c$, Superconductivity, Glass Model,  
Critical Current, Bean Model}

\vglue 0.3cm

\section{Introduction}

The high-$T_c$ glass model was introduced in 1987 \cite{MOR87,MOR89/2} to 
describe superconductivity in the high-$T_c$ cuprates (HTSC) \cite{BED86}. 
Originally the glass model was designed for s-wave symmetry of the superconducting
wave functions \cite{MOR87}. Whereas it was first argued \cite{SIG92}, that
the glass model is not applicable for d-wave symmetry, the experimental result 
of striped domains \cite{CHE89,TSU98} in the superconducting
CuO-planes inside the high-$T_c$ materials only gives rise to weak disorder. Thus 
the high-$T_c$ glass model is applicable in this situation, too \cite{MOR98}.
It was demonstrated \cite{MOR98}, that the 
high-$T_c$ glass model including the tt'--Hubbard model as a microscopic 
description of the striped superconducting domains is able to 
explain e.g.
the d-wave symmetry of the superconducting phase \cite{TSU94,KIR95/2}
and the pseudogap above $T_c$ in the density of states (DOS) \cite{DIN96,LOE96}.
In the combined high-$T_c$ glass and striped Hubbard model picture the stripes in the HTSC
occur at least at the same temperature, at which the pseudogap in the DOS opens,
because otherwise the striped Hubbard clusters, which are responsable for these gaps,
do not exist.

In this paper we will show, that within the high-$T_c$ glass model the 
maximum critical currents found in high-$T_c$ materials and their almost perfect
isotropy in a-b-direction can be understood. 
Following two independent paths  we calculate in the next section an upper bound 
for the critical current density $j_c$:
first with a direct calculation from the high-$T_c$ glass model \cite{EBN85}
and second considering the extended Bean model \cite{GYO89}. 
Afterwards we offer an intuitive picture 
for the (observed) isotropy of the critical currents and explain, why the
values of $j_c$ usually measured in high-$T_c$ materials are smaller than these 
upper bounds.  From these simple ideas follows 
a fabrication procedure, which could lead to a possible increase of $j_c$ 
in the high-$T_c$ materials.

\section{Critical Currents in the High-$T_c$ glass model}

\begin{figure}[hbtp]
\begin{center}
\begin{minipage}{5.0cm}

\epsfxsize 5.0cm \epsfbox{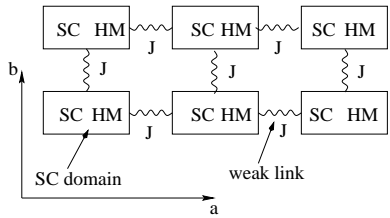}
\end{minipage}
\vglue0.2cm
\caption{\label{fig_stripes}
Superconducting (SC) domains (stripes) in the CuO-planes described
microscopically with the Hubbard model (HM).
}
\end{center}
\end{figure}

\begin{figure}[hbtp]
\begin{center}
\begin{minipage}{5.0cm}

\epsfxsize 5.0cm \epsfbox{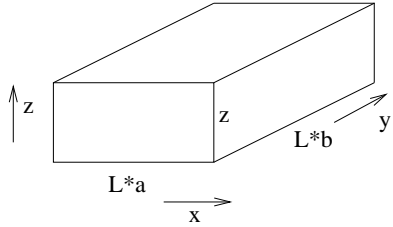}
\end{minipage}
\vglue0.2cm
\caption{\label{fig_geo}
Geometry and size of a single striped domain in the high-$T_c$ glass model.
}
\end{center}
\end{figure}

In the high-$T_c$ glass model a single superconducting CuO-plane is described as 
an array of striped domains, figure~\ref{fig_stripes}, 
\cite{MOR98}.
The typical dimension of a single striped domain with constant superconducting
phase inside the high-$T_c$ glass model can be roughly estimated
following e.\,g.\ Tsuei and Doderer \cite{TSU98} and Tranquada 
et.\ al.\ \cite{TRA97}.  We take for our calculation the values 
$a \approx 100$\,\AA, $b \approx 10$\,\AA, 
(e.g.\ $b \approx 16$\,\AA\ in LaSrCuO \cite{SAI97} or 
$b \approx 23$\,\AA\ in YBaCuO \cite{TSU98})
and $z \approx 10$\,\AA\ with the directions of figure~\ref{fig_geo}.
This choice of $a$, $b$, and $z$ is not crucial for our considerations, where
only their order of magnitude is important for the conclusions. 

In the following calculation we show that the glass-model leads
to critical current densities, which are in agreement with the experiments.
The starting point is the Hamiltonian for the glass model \cite{MOR87,MOR89/2}:
\[
\label{eqglass_h}
{\cal H} = - J \sum_{\left< i,k \right>} 
\cos \left(\phi_i - \phi_k - A_{i,k} \right)
\quad .
\]
The phase factors $A_{i,k}$ are given by
\[
A_{i,k} = \frac{2\pi}{\Phi_0} \int_i^j \vec{A} \, {\rm d} \vec{l}
\]
with ${\rm rot} \vec{A} = H \cdot \hat{z}$, where $\hat{z}$ is the 
unity vector in z-direction, and with $A_{i,k} = a_{i,k} \cdot H$ we get
\[
\label{eq_aij}
a_{i,k} = \frac{2\pi }{\Phi_0} \frac{x_i+x_k}{2} 
\left(y_k -y_i \right)
\quad .
\] 
In equation (\ref{eqglass_h}) up to  (\ref{eq_aij}) $\Phi_0$ 
is the elementary flux quantum, $\phi_i$ are the phases of the superconducting
wave functions, $J$ is the coupling energy between two clusters, 
$\left< i,k \right>$ is the sum over all nearest neighbors,  
and $x_i$, $x_k$, $y_i$, and $y_k$ are the coordinates of the 
center of gravity of the domains $i$ respectively $k$. 
Finally $H$ is the external magnetic field and $\vec{A}$ the corresponding vector
potential \cite{MOR87}. 

In the high-$T_c$ glass model only weak, ''correlated'' disorder was
chosen in the framework of the square lattice \cite{MOR87}.
We should note, that this weak disorder in the high-$T_c$ glass model can be described
as $x_i \approx i \cdot a$  and  $y_i \approx i \cdot b$ 
with the lattice constants $a$ in x-direction and $b$ correspondingly 
in y-direction \cite{MOR87,MOR98}.

To obtain the critical current density $j_{i,k}$ for a connection between
domain $i$ and $k$ we consider the Maxwell equation
\[
\label{eqmaxwell}
{\rm rot} \, H_{i,k} =  j_{i,k}
\quad .
\]
Next we assume a constant increase of the internal magnetic field analogous to 
the assumptions of the Bean model \cite{BEA62,KIM63,BEA64}:
\[
\label{eqbeanassumption}
{\rm rot} \, H_{i,k} = \frac{H_{i,k}}{\frac{x_i+x_k}{2}}
\quad .
\]

The free energy $F$ is given as $F=- k_B T \ln Z$, where the partition function
$Z = \sum_{\{ \Phi \} } \exp \left( - \beta {\cal H} (\Phi ) \right)$ with the inverse 
temperature $\beta \equiv 1/(k_B T)$ and the sum $\{ \Phi \}$ is over all possible configurations 
of the phases $\Phi=(\phi_1,\ldots,\phi_{L^2})$ for a lattice with $L\times L$ sites 
($k_B$ is the Boltzmann constant).
The magnetization $M$ of a sample is given by
$M = - 1/V  \cdot \partial F/\partial H$,
where $V$ is the volume of the sample. For the external
magnetic field $H=0$ we obtain:
\[
\label{eqmagnet_1}
M = \frac{1}{V} \frac{2\pi}{\Phi_0} J \sum_{\left<i,k \right>}
\frac{x_i+x_k}{2} \left( y_k -y_i \right)
\left< \sin \left( \phi_i - \phi_k \right) \right>
\]
where $\left< \ldots \right>$ dennotes the thermal expectation value.

The magnetization can be expressed in terms of the internal magnetic 
fields $H_{i,k}$
\[
\label{eqmagnet_2}
M = \frac{1}{2L^2} \sum_{\left< i,k \right>} H_{i,k}
\quad .
\]
Inserting equation (\ref{eqmagnet_1}) into equation (\ref{eqmagnet_2}),
using the abbreviation $I = J 2 \pi / \Phi_0$ and the volume $V = L a \cdot L b \cdot z$ 
(geometries of a striped domain, figure~\ref{fig_geo}) it follows 
\[
\label{eqhij}
H_{i,k} = \frac{1}{abz} 2 I \frac{x_i+x_k}{2}
\left( y_k - y_i \right)
\left< \sin \left(\phi_i - \phi_k \right) \right>
\quad .
\]

Using the equations (\ref{eqmaxwell}), (\ref{eqbeanassumption}), and (\ref{eqhij}) 
we get for the critical current density of the domains $i$ and $k$:
\[
j_{i,k} = \frac{I}{az} \left< \sin \left(\phi_i - \phi_k \right) \right>
\quad .
\]
Here we made also use of $y_k - y_i \approx b$, if $i$ and $k$ are nearest neighbors 
(n.n.) in y-direction and $y_k - y_i \approx 0$ in the other cases. 

Now we average over all $2L^2$ bonds between n.n.\ in the lattice and introduce 
$j_0 \equiv I / az$, where
$az$ is the area "used" by a single junction between two domains. We obtain:
\[
j = \frac{1}{2L^2} \sum_{\left< i,k \right>} j_{i,k}
= j_0 \frac{1}{2L^2} \sum_{\left< i,k \right>} 
\left< \sin \left( \phi_i - \phi_k \right)\right>
\quad .
\]
The average of the sinus:
\[
\label{eqbarsin}
\overline{\left< \sin \left( \phi_i - \phi_k \right)\right>} = 
\frac{1}{2L^2} \sum_{\left< i,k \right>}
\left< \sin \left( \phi_i - \phi_k \right)\right> 
\]
can be obtained from the simulation (index $_{sim}$) with
\[
\label{eqmsim}
M_{sim} = \frac{1}{2L^2} \sum_{\left< i,k \right>} 
\frac{x_{i,sim}+x_{k,sim}}{2}
\left( y_{k,sim} -y_{i,sim} \right)
\left< \sin \left( \phi_i - \phi_k \right)\right>
\quad .
\]
We now make following approximations: $y_{k,sim} - y_{i,sim} = 1$, if $i$ and $k$ are 
n.n.\ in y-direction, $y_{k,sim} - y_{i,sim} = 0$, if $i$ and $k$ are n.n.\
in x-direction, and 
$x_{i,sim} + x_{k,sim} \approx 2i$ neglecting the random placement of
the domains in the glass model \cite{MOR87}.

\begin{figure}[hbtp]
\begin{center}
\begin{minipage}{5.0cm}

\epsfxsize 5.0cm \epsfbox{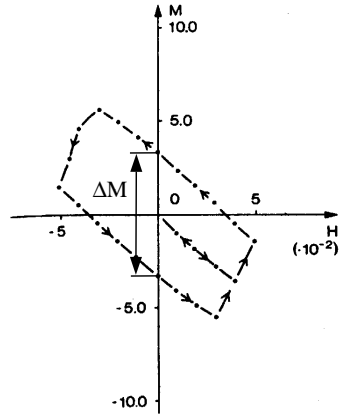}
\end{minipage}
\vglue0.2cm
\caption{\label{fig_magn}
Hysteresis measurement of the high-$T_c$ glass model. 
(analogous to figure 12 in $[$1$]$)
}
\end{center}
\end{figure}

With $i=0,1,\ldots,L-1$ this leads to the magnetization
\begin{eqnarray}
\label{eqbarsinmagn}
M_{sim} & = & 
\frac{L-1}{2} 
\overline{\left< \sin \left( \phi_i - \phi_k \right)\right>}
\end{eqnarray}
With $M_{sim} = 0.5 \Delta M_{sim} $ from figure~\ref{fig_magn} 
it follows
\[
\overline{\left< \sin \left( \phi_i - \phi_k \right)\right>} = \frac{1}{L-1} \Delta M_{sim}
\quad .
\]
Here also as in the Bean model $j \to j_c$ leads to the critical state and the critical current 
density is given by:
\[
\label{eq_jca_final}
j_c = j_0 \frac{1}{L-1} \Delta M_{sim}
\] 
with $j_0 = 2\pi J / (az\Phi_0) $.

\begin{figure}[hbtp]
\begin{center}
\begin{minipage}{5.0cm}

\epsfxsize 5.0cm \epsfbox{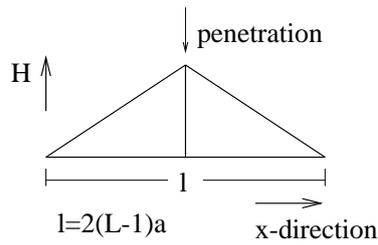}
\end{minipage}
\vglue0.2cm
\caption{\label{fig_pene}
Critical state in a single striped domain in the high-$T-c$ glass model.
}
\end{center}
\end{figure}

Now we calculate the critical current density in a second way.
Applying Bean's formula \cite{BEA62,BEA64} in the anisotropic case \cite{GYO89}
is justified as we have periodic boundary conditions in y-direction leading 
to a very long sample in y-direction. 
Concerning the internal magnetic fields in the critical state we have the situation 
illustrated in figure~\ref{fig_pene}. 
Therefore we have in the simulation in x-direction a sample size of double length. 
Thus the experimentally found magnetization $M_{exp}$ is the magnetization corresponding
to the triangle of figure~\ref{fig_pene}.

With the Bean formula of the anisotropic case \cite{GYO89}: 
\[
M_{exp} = \frac{j_c l}{20}
\]
and therefore with the length of the sample 
$l= 2(L-1)a$ (figure~\ref{fig_pene}) we have 
\[
\label{eqbean_exp}
j_c = \frac{20 M_{exp}}{2 (L-1) a}
\quad .
\]

Now we calculate the value of the ''experimental'' magnetization $M_{exp}$ from $M_{sim}$:
\[
M_{exp} = \frac{1}{V} \frac{2\pi}{\Phi_0} J 
\sum_{\left< i,k \right>} \frac{x_i + x_k }{2} \left( y_k-y_i \right)
\left< \sin \left( \phi_i - \phi_k \right)\right>
\]
with $x_i = a \cdot x_{i,sim}$ ($x_{i,sim} = 0, 1, L-1$) and  $y_i =  b \cdot y_{i,sim}$ 
($y_{i,sim} = 0, 1, L-1$) we have
\[
\label{eqm_exp_sim}
M_{exp} = 
\frac{1}{L^2 abz}  \frac{2\pi}{\Phi_0} J 
ab   
\sum_{\left< i,k \right>} \frac{x_{i,sim} + x_{j,sim} }{2} \left( y_{j,sim} - y_{i,sim} \right)
\left< \sin \left( \phi_i - \phi_k \right)\right>
\quad .
\]
Therefore with $I$ from equation (\ref{eqhij}) and $M_{sim}$ from equation (\ref{eqmsim}) 
and with $ \Delta M_{sim} = 2 M_{sim}$ (figure~\ref{fig_magn}) equation 
(\ref{eqm_exp_sim}) leads to
\[
M_{exp} = 
\frac{1}{z} I \Delta M_{sim}
\quad ,
\]
which leads (with equation (\ref{eqbean_exp})) to the critical current density
\[
j_c =   \frac{I}{az} \frac{10}{L-1} \Delta M_{sim}
\quad .
\]

Using the abbreviation $j_0 = I/(az)$ and measuring $j_c$ in A/cm$^2$  
\cite{BEA62} we have finally
\[
\label{eqjc_bean}
j_c = j_0 \frac{1}{L-1} \Delta M_{sim}
\]
as in our first calculation (equation (\ref{eq_jca_final})) of the critical current 
density using Bean's assumption 
(equation (\ref{eqbeanassumption})) in the high-$T_c$ glass model directly.

Using the numerical values $\Delta M_{sim} \approx 6$ and $L-1 = 15$ from figure~\ref{fig_magn} 
and the experimental values $a = 100\,{\rm \AA} = 10^{-8} \,{\rm m}$ and 
$z= 10\, {\rm \AA} = 10^{-9}\, {\rm m}$
for the stripes in the HTSC \cite{SAI97,TSU98} we get $I \approx 4.2 \cdot 10^{-6} {\rm A}$. 
And therefore we have for the critical current density
\begin{eqnarray}
\label{eq_jc_glass_wert}
j_c & = & \frac{I}{az} \Delta M_{sim} 
\approx 1.7 \cdot 10^{7} \frac{{\rm A}}{{\rm cm}^2}
\quad ,
\end{eqnarray}
which is surprisingly close to the experimental values 
$j_c \approx 5 \cdot 10^7{\rm A}/{\rm cm}^2$  at 4\,K
and zero field for the best films \cite{BUR92} and higher than $j_c$
in wires ($j_c < 10^6$\,Acm$^{-2}$, e.\,g.\ \cite{HAY96,SEE98}). We want to note, that in figure~\ref{fig_magn} 
the $\Delta M_{sim}$ was measured at $T=0.2J$ which corresponds to $20$\,K for 
$T_c= 100$\,K.

To calculate critical current densities with equation (\ref{eq_jca_final}) resp.\ (\ref{eqjc_bean}) 
we made use of 
Bean's assumption of a constant increase of the internal magnetic fields $H_{i,k}$. It is 
clear, that this is a rather crude approximation for the high-$T_c$ glass model. 
Furthermore for the calculation of $j_c$ we used the approximation of the averaged sinus 
(equation (\ref{eqbarsin})). This gives perhaps a wrong estimate of the true critical 
current density of the high-$T_c$ glass model. 

To support our line of reasoning we turn back to the glass model and
estimate $j_c$ in the glass model directly following 
Ebner and Stroud \cite{EBN85}.
From the knowledge of the size of the domains it is possible to obtain 
the critical current $J_c$ of the glass model, which can be determined directly
\cite{AMB63,SHI84,EBN85}
\[
\label{eqjc_glass_2}
J_c = \frac{2e}{h} J
\]  
with $J= k_B T_c$ \cite{MOR87} and $T_c = 100$\,K. ($k_B$ is the Boltzmann constant,
$e$ the elementary electron charge and $h$ the Planck constant.) 
Equation (\ref{eqjc_glass_2}) gives an upper bound for the maximum current 
through a single Josephson junction in the high-$T_c$ glass model \cite{AMB63}, which is
used to estimate the coupling energy $J$ in the high-$T_c$ glass model \cite{SHI84}.  
To determine the critical current density
$j_c$ in a-direction we consider in a single domain (in figure~\ref{fig_stripes}) the area $A$ in 
b-z-direction, through which the current flows, figure~\ref{fig_geo}. With 
$A \approx b \cdot z \approx  10^{-18}$\,m$^2$ 
we obtain the critical current density 
\[
\label{eqjc_glass_3}
j_c = \frac{J_c}{A}  = \frac{2e}{h} \frac{k_B T_c}{bz} 
= 0.667 \cdot 10^8 \frac{{\rm A}}{{\rm cm}^2}
\quad .
\]
This is indeed an upper bound to $j_c$ in equation (\ref{eq_jc_glass_wert}).
Thus both calculations (equation (\ref{eq_jca_final}) and (\ref{eqjc_bean})) 
of $j_c$ give the same formula for $j_c$, which is lower than the 
upper bound for the high-$T_c$ glass model in equation (\ref{eqjc_glass_3}).

\section{(An)isotropy of the critical current}

Next we consider the critical currents in the b-direction instead of the 
a-direction.
First we determine the magnetization (with $I=(2\pi/\Phi_0)J$)
\[
M = \frac{I}{z} \Delta M_{sim} \approx 260 \frac{{\rm A}}{{\rm cm}}
\quad ,
\]
which is in good agreement with the experiments \cite{GYO89}, too.
Note, that $M$ is independent of the size of the domains in the
a-b-plane.
Repeating the above calculations we have for the critical current density in b-direction: 
\[
j_c = \tilde{j}_0 \frac{1}{L-1} \Delta M_{sim} 
\quad .
\]
This is analogous to equation (\ref{eq_jca_final})
with a different $\tilde{j}_0 = I/(bz)$, in which $a$ is replaced by $b$.
Inserting the experimental values $b \approx a/10 = 10$\,\AA\ and
$z \approx 10$\,\AA\ \cite{SAI97,TSU98} we obtain
\[
\label{eq_jcb}
j_c^b  \approx 10 \cdot j_c^a = 1.7 \cdot 10^8 \frac{{\rm A}}{{\rm cm}^2}
\quad . 
\]
In b-direction the upper bound analogous to equation (\ref{eqjc_glass_3})
also exists. Of course this upper bound is now also about ten times larger 
than in a-direction.
 
Thus the consideration of stripes in the high-$T_c$ glass model leads to a 
strong anisotropy of about the factor 
ten for the critical current densities between a- and b-direction depending on 
the ratio $a/b$ of the striped domains. But this 
relatively large factor was never reported in the experimental literature.

We want to note, that on the one hand the size of the stripes enters
the calculation of $j_c$ reciprocally, but on the other hand the number of 
domains in one spatial direction influences $j_c$, too. Thus a better
knowledge of the size of the domains is desirable. Additionally a 
finite size scaling (or the simulation of different system sizes $L$)
for the magnetization $\Delta M_{sim}$ of the high-$T_c$ glass model 
is necessary to calculate the critical current densities more accurately.

Now we have two features (anisotropy of $j_c$ in a- and b-direction and a
$j_c$, which is in the order of magnitude of the largest experimentally 
found $j_c$ in HTSC) of the high-$T_c$ glass model, which do not agree with the
experiment. In our opinion there are two possible mechanisms, which can
lead to a small or vanishing anisotropy.  The first one is an anisotropic
coupling constant $J$ ($J_a \ne J_b$) in the high-$T_c$ glass model, which 
may be different in a- and b-direction. But this probably only reduces the 
anisotropy. And it is unlikely, that these anisotropic coupling constants will
lead to an isotropic critical current density $j_c$.

\begin{figure}[hbtp]
\begin{center}
\begin{minipage}{7.0cm}

\epsfxsize 7.0cm \epsfbox{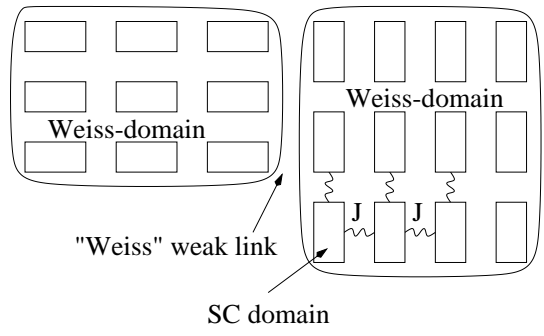}
\end{minipage}
\vglue0.2cm
\caption{\label{fig_weiss}
Weak links between Weiss-domains.
}
\end{center}
\end{figure}

We postulate therefore the existence of Weiss-type domains in the 
planes with the stripes, which are dominantly either pointing in 
a- or b-direction, figure~\ref{fig_weiss}. 
The existence of the Weiss-domains on the other hand explains the relatively
lower critical current densities $j_c$ found in experiments for single crystals 
and thin films and in particular their differences.

The resulting weak links between the Weiss-domains (figure~\ref{fig_weiss}),
which have to be assumed to be (much) "weaker", than the weak links of the 
high-$T_c$ glass model, restrict $j_c$ to lower values. Thus equation
(\ref{eq_jc_glass_wert}) and (\ref{eq_jcb}) are only upper bounds of $j_c$, too.

This possibility was first brought to our attention in private discussions with
Tsuei and Doderer. In the light of the above estimated factor of $a/b \approx 10 $ 
this picture is in our view one possible explanation for the experimental situation. 
Crystals or thin films with a predominant direction of the stripes 
are therefore predicted to show this anisotropy. 
But the main problem are the weak links between the Weiss-domains. 
These obviously govern the final experimental measurement
of the critical current density. This picture could also explain the differences  from sample to 
sample. While $T_c$ is for all samples (nearly) identical, $j_c$ is quite different.
But in our theoretical calculations for the high-$T_c$ glass model both $T_c$ and $j_c$
only depend on the coupling $J$. 

Considering this picture it is clear, that improvements of $j_c$ rise
in general the quality of the samples. Especially in the light of the possible
fabrication of electronic devices, better high-$T_c$ samples have to be obtained 
by removing the Weiss-domains and the weak links between these domains. 
Therefore it would be extremely fruitful to avoid the Weiss-domains or 
to restrict the influence of the resulting weak links. 
In this spirit we propose a receipt following from the fabrication of magnets. 
The magnets are cooled down in a strong magnetic field. For high-$T_c$-materials
we propose a similar procedure. The cooling process should use the cooling
schedule known from simulations. This schedule has been subject to extensive 
research in the field of (physical) optimization in particular the "simulated 
annealing" method \cite{KIR83,MOR87/2}.

Procedures developed in the optimization theory \cite{KIR83,MOR87/2,SCH96/4,SCH98}
may be directly transfered to the annealing of the high-$T_c$-materials. 
Furthermore the annealing has to be carried out in an electric field and/or 
while an electric current is flowing through the sample. 

We expect from these procedures a substantial increase of the quality of the 
high-$T_c$-materials. In particular the fabrication of electronic devises as e.g.\ 
transistors should benefit greatly from these ideas.
In the fabrication of wires or polycrystals additionally the weak links 
between the superconducting grains limit the critical current densities
and should be removed, too.

\section{Summary and Conclusions}

In the high-$T_c$ glass model the striped superconducting domains of the 
high-$T_c$ materials \cite{TSU98,TRA97}
are the domains of constant superconducting phases \cite{MOR98}.
The size of the striped domains may be deduced by mainly theoretical 
considerations and magnetic measurements \cite{MOR98}. 
These measurements and the corresponding theoretical framework were already known shortly 
after the discovery of the HTSC \cite{MOR87}. Already there the size of the 
domains could be estimated correctly to be approximately $10^4$\,\AA$^2$ \cite{MOR98}. 
This leads to the conclusion, that the 
glass behavior is not related to the early ceramic structure of the grains, that
many but not all samples exhibit. 

From the high-$T_c$ glass model the critical current density can be calculated
following two different paths: first using the extended Bean model \cite{BEA62,GYO89}
and hysteresis measurements of the high-$T_c$ glass model and second directly
from the definition of the glass model using Bean's assumption \cite{BEA62} of the 
critical state model. Both approaches lead consistently to the same 
critical current densities $j_c$ and
to a strong anisotropic critical current density for the a- resp.\ b-direction of $j_c$, which 
is much larger than the anisotropy found experimentally.
But they are close to the highest measured critical current densities in thin films.
This strong anisotropy follows from the underlying striped shape of the domains.

Taking the same type of approximation (following the Bean model) we obtain the same value 
of $j_c$ in the simulation as in the experiments. This is in our opinion a strong evidence 
in favor of the high-$T_c$ glass model.

Nonetheless it is puzzling, why this anisotropy in the a-b-plane was never 
reported for $j_c$. We propose two extensions of the high-$T_c$ glass model: first
an anisotropic coupling $J_a \ne J_b$, which in our opinion can only reduce the
anisotropy, and second the existence of Weiss-type domains, in which 
the predominant direction of the stripes is turned by 90$^\circ$. The latter leads to 
nearly isotropic critical current densities and lowers $j_c$ on account of the ''new'' 
weak links between  these areas with the same direction of the stripes. 
This explains, too, why 
$T_c$ is nearly constant between different samples whereas $j_c$ is quite different.
Removing these Weiss-type domains should lead to higher critical current densities $j_c$, 
which may be done by simulated annealing of the HTSC (eventual in electric fields).

\section{Acknowledgment}

We want to thank C.C. Tsuei and T. Doderer 
for very helpful discussions. We would like to acknowledge illuminating discussions with 
K.A. M{\"u}ller, H. Keller, T. Schneider,  E. Stoll, and J.M. Singer. 
Especially we would like to thank P.C. Pattnaik and D.M. Newns for inspiring discussions and ideas.
Finally we acknowledge the financial support of the UniOpt GmbH, Regensburg.



\end{document}